\documentclass[conference,a4paper]{IEEEtran}
\ifCLASSINFOpdf
\else
\fi
\usepackage{graphicx}
\usepackage{bm}
\usepackage{booktabs}
\usepackage{multirow}
\usepackage{stfloats}

\begin{document}
%
\title{EAR-NET: Error Attention Refining Network For Retinal Vessel Segmentation}

\author{\IEEEauthorblockN{Jun Wang$^1$, Yang Zhao$^{2,3}$,Linglong Qian$^4$,  Xiaohan Yu$^2$, Yongsheng Gao$^2$}
\IEEEauthorblockA{$^1$University of Warwick $^2$Griffith University $^3$University of Adelaide $^4$King's College London\\
jun.wang.3@warwick.ac.uk, linglong.qian@kcl.ac.uk, yang.zhao01@adelaide.edu.au \\xiaohan.yu@griffith.edu.au,  yongsheng.gao@griffith.edu.au}

}



%



\maketitle

\begin{abstract}
The precise detection of blood vessels in retinal images is crucial to the early diagnosis of the retinal vascular diseases, e.g., diabetic, hypertensive and solar retinopathies. Existing works often fail in predicting the abnormal areas, e.g, sudden brighter and darker areas and are inclined to predict a pixel to background due to the significant class imbalance, leading to high accuracy and specificity while low sensitivity. To that end, we propose a novel error attention refining network (ERA-Net) that is capable of learning and predicting the potential false predictions in a two-stage manner for effective retinal vessel segmentation. The proposed ERA-Net in the refine stage drives the model to focus on and refine the segmentation errors produced in the initial training stage. To achieve this, unlike most previous attention approaches that run in an unsupervised manner, we introduce a novel error attention mechanism which considers the differences between the ground truth and the initial segmentation masks as the ground truth to supervise the attention map learning. Experimental results demonstrate that our method achieves state-of-the-art performance on two common retinal blood vessel datasets.
\end{abstract}


%
\IEEEpeerreviewmaketitle

\section{Introduction}
Retinal examination is an important diagnostic method to certain pathological diseases such as diabetes, high blood pressure, hypertension. Retinopathies may deteriorate into blindness or lead to the loss of vision. Fortunately, these severe situations usually can be averted by the timely screening and treatment \cite{shoemaker2002vision}. However, this is a time-consuming task since clinical doctors with professional expertise are required to examine considerable retinas. In addition, false diagnosis can be made due to some subjective factors, e.g, fatigue of clinical experts. Moreover, clinical experts are limited in some countries, especially in developing countries, hence humans living in these countries have difficulties in accessing the medical resources and receiving the treatment for the retinopathies in time.
\par
To alleviate the shortage of the medical resources, computer algorithms \cite{chen2018s3d,fu2016deepvessel,gu2019net,guo2021sa,hoover2000locating,irshad2014classification,jin2019dunet,kande2010unsupervised,lam2010general,li2020iternet,li2018h,martinez2007segmentation,mou2019cs,palomera2009parallel,ronneberger2015u,salazar2010retinal,shin2019deep,wang2018interactive,zhang2019net,zhou2018unet++} have been developed and introduced to automatic retinal image analysis. The shape of the blood retinal vessels has been used as a good indicator to detect the retinal vein occlusion \cite{zana1999multimodal}, grade the tortuosity for hypertension \cite{irshad2014classification}, and diagnose the glaucoma \cite{youssif2007optic} and diabetic retinopathy \cite{smart2015study}. The blood vessels segmentation is the first key step for detecting the eye-related diseases. However, this is a quite challenging task and far from being solved mainly due to the super-complicated structures of the retinal blood vessels, subtle differences in appearance of vessels from the background, possible improper illumination and sensor noises \cite{li2020iternet}. Despite significant improvement achieved in recent computer-aided methods, there are some problems causing considerable false predictions, limiting their performance in real world applications. One major problem in the deep learning based models defecting the retinal blood vessel segmentation performance is that they normally recover the high-resolution representation from the low-resolution features. Essential information may be lost during this phase, causing inaccurate results. Besides, most existing methods struggle to handle the pixels around the brighter and darker spots as shown in the rectangulars in Figure 1, leading to the false predictions \cite{li2020iternet}. Moreover, most of exiting works are dominated by the background due to the significant imbalance between the foreground and background, leading to high accuracy and specificity while low sensitivity.\par
To cope with these problems, we propose EAR-Net, an error attention refining network for retinal vessel segmentation. The proposed EAR-Net refines initial segmentation results produced by the first-stage trained segmentation model, thus explicitly driving the refinement model to focus more on the false results in the initial segmentation results. We develop an error attention approach to enable the EAR-Net to localize the possible false prediction. Different from most existing attention mechanisms that are more likely to be trapped into the local optimum due to their unsupervised scenario, our proposed error attention employs the initial predicted masks as the auxiliary supervision signal to aid the learning of the error attention, hence alleviating the local optimum problem. A visual illustration of the segmentation results obtained by the proposed EAR-Net is shown in Fig. 2. \par
\begin{figure}[t]
\centering
 \includegraphics[width=0.5\textwidth]{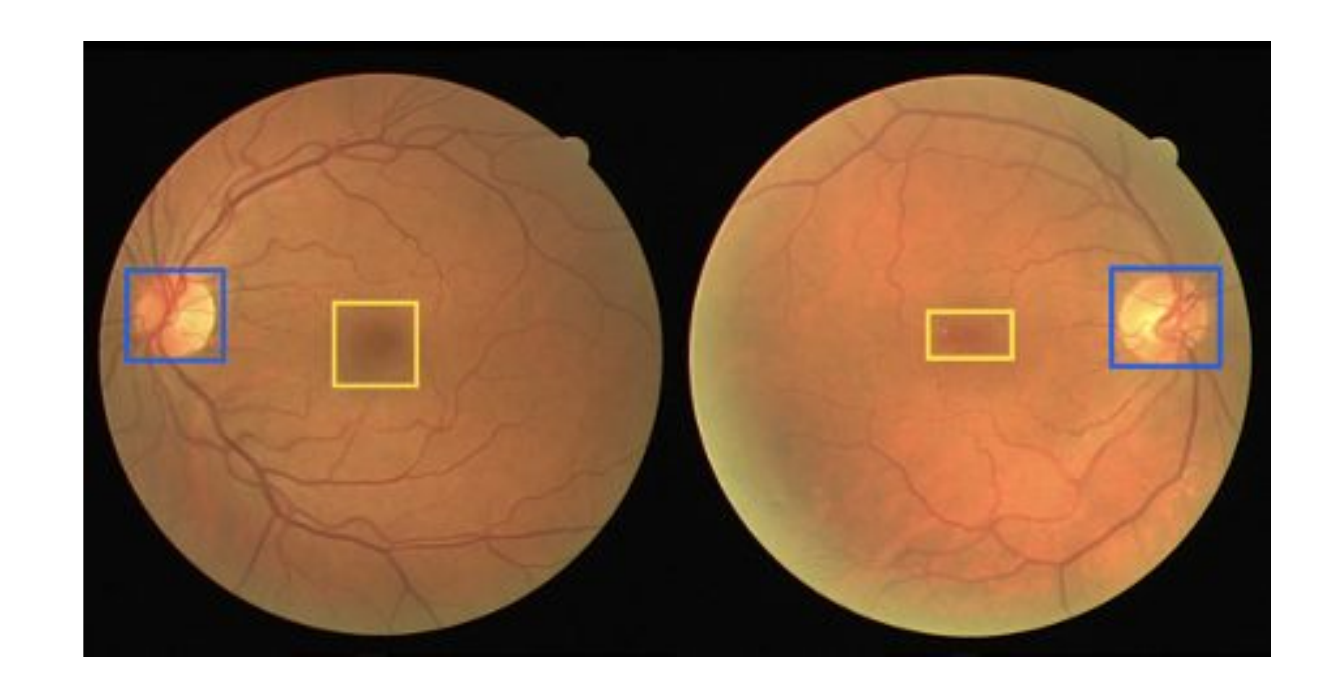}
 \caption{An illustration to show the brighter and darker areas in retinal images. The brighter areas are demonstrated in the blue rectangular, while the darker areas are shown in the yellow rectangular.} \label{fig1} 
 \end{figure}

This work mainly has three contributions. (1) We propose a novel error attention refining network (ERA-Net) that is capable of learning and predicting the potential false predictions in a two-stage manner for effective retinal vessel segmentation.  (2) We develop an error attention mechanism to enable the false prediction localization capability of ERA-Net, by employing the initial predicted masks as auxiliary supervision signal to aid the learning of the error attention; (3) Experimental results demonstrate that our method achieves state-of-the-art performance on the DRIVE \cite{staal2004ridge} and STARE  \cite{irshad2014classification} datasets.

\begin{figure}[t]
\begin{center}
\includegraphics[width=0.45\textwidth]{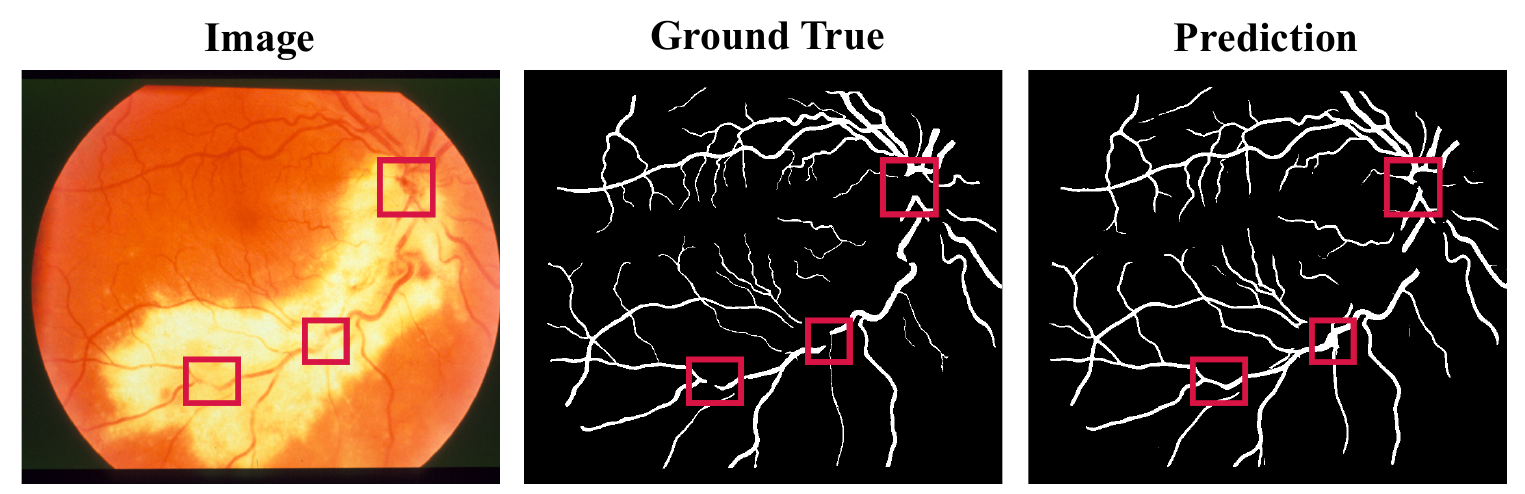}
\caption{A visual illustration of retinal blood segmentation result from the test set of STARE dataset by our proposed method.} \label{fig2}
\end{center}
\end{figure}
\section{Related Works}
Existing approaches on retinal blood vessels segmentation can be coarsely divided into two groups: unsupervised methods and supervised methods. Following sub-sections briefly introduce some representative works from these two groups. 
\subsection{Unsupervised methods}
There is no training phase in unsupervised methods. Martinez-Perez et al.\cite{martinez2007segmentation} proposed to 
segment the retinal blood vessels via multi-scale feature extraction. They took advantage of the first and second derivatives of the intensity images to handle the large variations in contrast in retinal images. Salazar-Gonzalez et al. \cite{salazar2010retinal} suggested to perform rough segmentation from the augmented images and then construct a graph based on the rough segmentation results to obtain the final results. Yavuz et al. \cite{yavuz2011retinal} used Gabor filter to detect the blood vessels and exploited a top-hat transformation to improve the performance. Palomera-Perez et al. \cite{palomera2009parallel} presented an efficient segmentation algorithm by partitioning the images and processing the sub-images in parallel. 
These methods usually make use of the hand-crafted and local features and have gained potent results. However, they lacks robustness and generalization capability, suffering severely from the abnormal pixels such as brighter and darker areas as illustrated in Figure 1. \par 
\subsection{Supervised methods}
Supervised methods contains a training phase based on the manually labelled annotations. Many existing state-of-the-art approaches on retinal image segmentation are based on the UNet architecture due to its excellent performance on medical image segmentation. Jin et al. \cite{jin2019dunet} proposed a DUNet model which utilized the deformable convolutions to enlarged the receptive field based on the vessels' scales and shapes. Guo et al. \cite{guo2021sa} developed a spatial attention mechanism, the shape attention, to refine the feature maps and demonstrated improved results. A number of non-UNet based models also achieved great success. Shin et al. \cite{shin2019deep} argued that graphical structure of vessels shape contributes positively to the segmentation accuracy. They utilized both the local appearance and the neighborhood relationships to segment retinal blood vessels by integrating a graph neural network (GNN) to a convolutional neural network (CNN). Although these deep learning based methods have demonstrated encouraging results on retinal image segmentation task, limitation exists as stated in the previous section. To mitigate these problems, we propose an error attention based refinement approach which drives the model to pay attention to the false predictions in the initial segmentation results. 
This is also in accordance with the spirit of recent research \cite{wang2021mask,wang2021feature,wang2021boosted, YUICCV21,YU2021108067,ZHAO2021107938,yu2020patchy,wang2021boosted,zhao2022learning,yu2019multiscale,yu2015leaf,yu2016multiscale}, which focuses on localizing subtle yet vital regions.

\section{Methods}
This section elaborately demonstrates our proposed method, the error attention based refinement. The overall architecture is shown in Figure 2. It is a two-stage training. In the first stage, an image segmentation model is trained based on the training samples. Then, the training images are fed into trained model to obtain the initial segmentation results. In the second stage, combined with the ground truth, the initial predicted masks are sent to the error attention module to produce the error attention maps. The error attention module and the backbone are trained jointly and end-to-end. It should be noted that the first stage and the second stage models share the same trunk model. The second stage model is fine-tuned according to the training samples and the error maps. The semantic logits are refined by aggregating the original semantic logits and the learned error attention maps.  

\begin{figure*}[htbp]
\centering
\includegraphics[width=\textwidth]{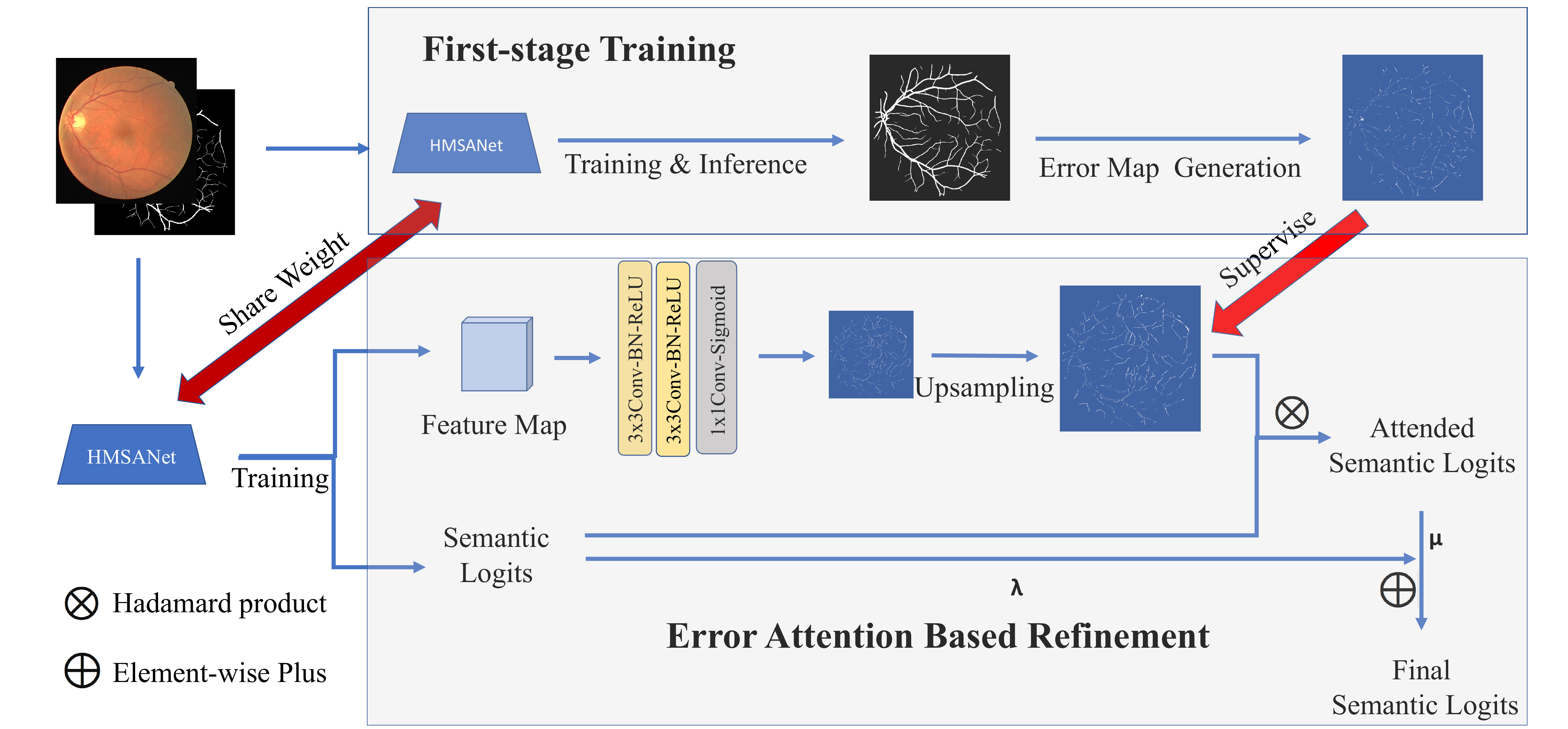}
\caption{The overall architecture of our proposed method. It comprises two-stage training. The first stage trains a segmentation model and generate the initial segmentation results which will then be refined by aggregating the attended semantic logits and the original semantic logits in the refinement phase. The attended semantic logits is obtained via the error attention module which is supervised by the error maps.} \label{fig1}
\end{figure*}

\subsection{Initial Retinal Blood Vessels Segmentation}
To better interpret our proposed method, we first briefly introduce the HMSANet \cite{tao2020hierarchical}. We denote a training image and its ground truth as $<\bm{I,GT}>$, and use $r$ to denote the image scale factor. Given two image scale factors $r=\bm{\alpha}$ and $r=\bm{\beta}$, the corresponding two scales of images $\bm{I_{r=\alpha}}$ and $\bm{I_{r=\beta}}$ are generated and then fed into the backbone HRNet-OCR \cite{yuan2020object} to obtain the semantic logits $\bm{l_{r=\alpha}}$ and $\bm{l_{r=\beta}}$. Given $\bm{\alpha}=0.5$ and $\bm{\beta}=1$, the final semantic logits of the original scales (1x) $\bm{l_{r=1}}$ are then calculated by:
 \begin{equation}
  \bm{l_{r=1}}= \bm{U}(\bm{l_{r=0.5}}*\bm{a_{r=0.5}})+(1-\bm{U}(\bm{a_{r=0.5}}))*\bm{l_{r=1}}
 \end{equation}

 \noindent
 Where $\bm{U}$ is the upsampling operator and $*$ is the Hadamard product. $\bm{a}$ denotes the scale attention maps.

 Different from most existing scale attention works that learn attention maps for every member in a fixed set of scales, HMSANet learns the relative attention between adjacent scales \cite{tao2020hierarchical}. The model is trained only with the adjacent image pairs, but can efficiently and effectively performs hierarchical scales augmentation method during the inference. For instance, three scales, 0.5x, 1x and 2x, data augmentation method is applied during the inference. Nonetheless, HMSANet only learns the relative attention between 0.5x and 1x images in the training phase. This relative attention then can be simply applied to the 1.0x and 2.0x images during the inference, and no extra training procedure is needed.\par
We adopt the HMSANet as the segmentation baseline network due to its state-of-the-art performance on image segmentation. As mentioned in the previous section, HMSANet preserves the high-resolution during the whole training processes, hence essential information is more likely to be maintained, motivating us to select it as our backbone rather than UNet-based architecture. \par

\subsection{Error Attention Module}
Error attention module aims to recognize the potential errors of the initial segmentation results and drive the refinement-stage training to focus on these false predictions. The bottom rectangular in Figure 2 illustrates the Error Attention Module (EAM). The EAM is added on top of the baseline on where the output features are four times smaller than the original image size. We implement our ideas via the following  three steps. \par
\subsubsection{Error Maps Generation} Firstly, we should obtain the error maps which are then considered as the ground truth of the error attention maps. To achieve this, training images are sent to the first-stage trained model again to obtain their predicted masks denoted as $\bm{M}_1$. It should be mentioned that the prediction result for a pixel is in the set \{0,1\}, where 0 is for the background and 1 indicates the presence of the blood vessel. Given the initial segmentation masks $\bm{M}_1$ and the ground truth $\bm{GT}$, the error maps $\bm{Em}$ are calculated by:

\begin{equation}
\bm{Em}_{(i,j)}=\left
\{
             \begin{array}{ll}
             1, \bm{GT}_{(i,j)} & > \bm{M}_{1,(i,j)}, \\
             0, \bm{GT}_{(i,j)} & \leq \bm{M}_{1,(i,j)}.
             \end{array}
\right.
\end{equation}

\noindent
where $i$ and $j$ are the spatial coordinates in the $\bm{EM}$, $\bm{GT}$, and $\bm{M}_1$.
The effectiveness for this design is that the prediction of a pixel is a false negative when the ground truth is larger than the prediction. Hence, we set the error map to one to enhance the activation in this point. Similarly, response should be reduced for a false positive point.

\subsubsection{Error Attention Subnetwork.}
Having obtained the error maps, we then present the proposed error attention subnetwork. The error attention subnetwork consists of three units as shown in the Figure 2. The first two units comprise a 3x3 Conv layer, a Batch Normalization layer and a ReLU activation layer to perform task-specific feature representation. The last unit utilizes an 1x1 Conv layer to map the number of channel to one. Finally, a Sigmoid function is employed to render the response to fall into the range of [0,1]. We denote the feature maps sent to the error attention subnetwork as $\bm{fm}\subseteq\bm{R}^{H\times W \times C}$. Note that $H$, $W$, $C$ are the height, width and the channel size, respectively. Then, this process is summarised in Equations (3) and (4).

\begin{equation}
 \bm{o}=\delta(BN(f_{3 \times 3}(\delta(BN(f_{3 \times 3}(\bm{fm}))))))
\end{equation}

\begin{equation}
 \bm{Am}=\sigma(f_{1 \times 1}(\bm{o}))
\end{equation}

\noindent
where $\bm{o}$ is the output of the second unit. $\bm{Am} \subseteq\bm{R}^{H\times W \times 1} $ is the final error attention map. $f_{* \times *}$ denotes the Conv layer with the kernel size of $*\times *$. $BN$, $\delta$ and $\sigma$ are the Batch Normalization layer, ReLU activation function and Sigmoid function, respectively. \par

\subsubsection{Error Attention Loss}
Finally, unlike most existing attention mechanisms that run in an unsupervised manner, the error attention module considers the error maps as the ground truth. Therefore, it is less likely to be trapped into the local optimum. The optimization of the error attention subnetwork is driven by a Mean Squared Error (MSE) Loss demonstrated in Eq (5).

\begin{equation}
 L_{EA}=\frac{1}{H\times W}\sum_{i=1}^{H}\sum_{j=1}^{W}||\bm{Em}_{i,j}-\bm{Am}_{i,j}||
\end{equation}
where $i$ and $j$ are the coordinates in the error maps $\bm{Em}$ and the attention maps $\bm{Am}$.\par

\subsection{Refinement}
After obtaining the error attention maps $\bm{Am}$, we refine the semantic logits $\bm{l}$ in the refinement-stage training. Predicting the possible errors is of great difficulty. Therefore, the error attention maps may contain a number of false predictions. To alleviate the influence of these false results, instead of directly regarding the attended semantic logits as the final semantic logits, we take advantage of both the original semantic logits $\bm{l}$ and the attended semantic logic $\bm{l}_a$ to generate the final semantic logits $\bm{l}_{final}$ as shown in Equations (6) to (7). 

 \begin{equation}
  \bm{l_a}=\bm{Am}*\bm{l}
 \end{equation}

\begin{equation}
  \bm{l}_{final}=\lambda\bm{l}+\mu\bm{l_a}
\end{equation}
where $\lambda$ and $\mu$ are two hyper-parameters to balance the contribution between two features, and $\lambda+\mu=1$. $*$ is the Hadamard product.

\subsection{Objective Functions}
In the first stage training, we follow the same objective functions of baseline which adopts the cross entropy loss $L_{CE}$ and a heatmap estimation loss $L_{HM}$ to train the model. An error attention loss $L_{EA}$ is added to supervise the learning of the error attention subnetwork in the refinement phase. Consequently, the final objective functions $L$ in the refinement phase are formed as:

\begin{equation}
  L=\eta L_{CE} + \gamma L_{HM} + \epsilon L_{EA}
\end{equation}
where $\eta$ and $\gamma$ and $\epsilon$ are three hyper-parameters to balance the contribution among the three objective functions.

\section{Experimental Results}
Following most existing works \cite{fu2016deepvessel,jin2019dunet,li2020iternet,ronneberger2015u}, we verify the effectiveness of the proposed method on two public retinal image benchmarks, the DRIVE \cite{staal2004ridge} database and the STARE \cite{hoover2000locating} database. Figures 3 and 4 visualize some examples from these two datasets. Since training set and test set are not explicitly split on STARE dataset, we follow the same setting of \cite{mou2019cs} which performs 4-fold cross validation. The reported results of STARE benchmark are the average values among all folds. Three common evaluation metrics shown in Equations (9)-(11) are used to assess the performance of the model.
\begin{equation}
  \bm{accuracy}(ACC)=\frac{TP+TN}{TP+TN+FP+FN}
\end{equation}

\begin{equation}
  \bm{sensitivity}(SE)=\frac{TP}{TP+FN}
\end{equation}

\begin{equation}
  \bm{specificity}(SP)=\frac{TN}{TN+FP}
\end{equation}

\begin{figure}[!h]
\begin{center}
\includegraphics[width=0.4\textwidth]{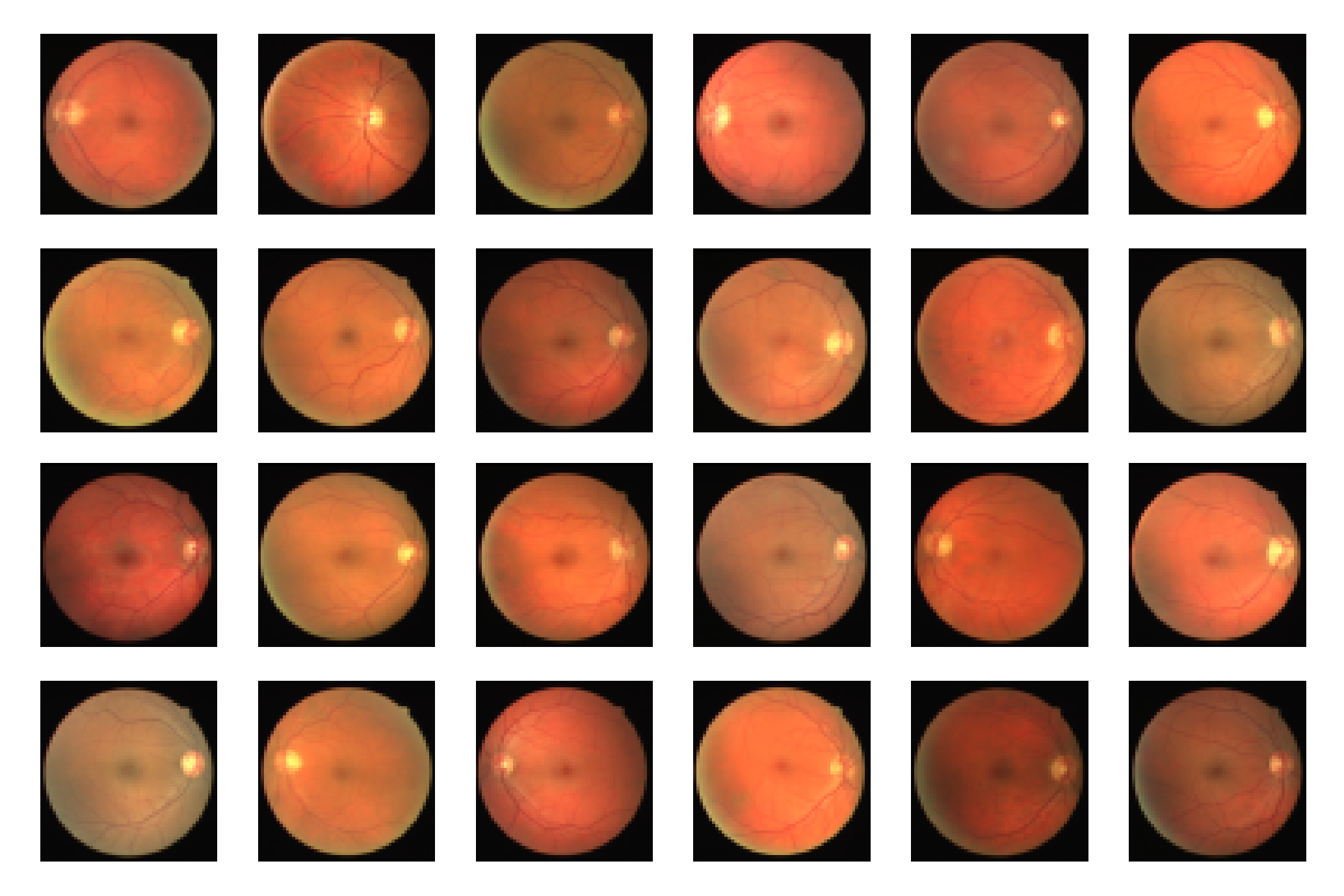}
\caption{24 examples from the DRIVE dataset.} \label{fig3}
\end{center}
\end{figure}

\begin{figure}[!h]
\begin{center}
\includegraphics[width=0.4\textwidth]{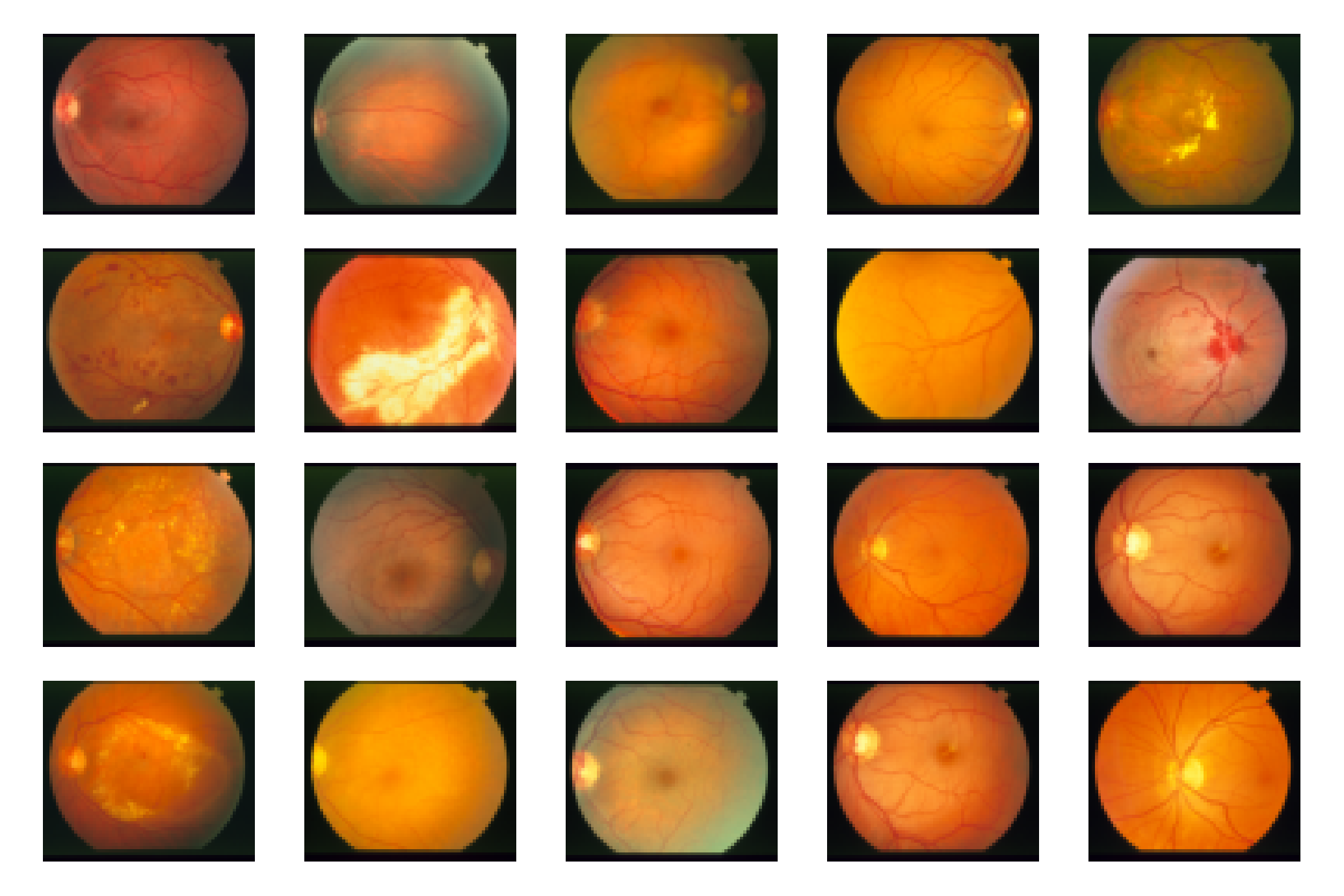}
\caption{18 examples from the STARE dataset.} \label{fig3}
\end{center}
\end{figure}

\subsection{Implementation Details}
We adopt the same data augmentation strategy of \cite{guo2021sa} to enrich the training set. $\lambda$, $\mu$, $\eta$, $\gamma$ and $\epsilon$ are set to 0.5, 0.5, 1, 0.4, 0.5, respectively. The stochastic gradient descent (SGD) optimizer is selected with a momentum of 0.9 to optimize the model. In the first stage, we train the model 50 and 40 epochs for the DRIVE and STARE datasets, respectively. The trained model are then refined by our proposed method for 15 epochs. The learning rate is set to 0.005 in the first-stage and 0.001 in the refinement phase with the learning rate decay.

\subsection{Segmentation Results on DRIVE and STARE}
Table 1 lists the segmentation results of the proposed EAR-Net and the state-of-the-art methods. We observe that EAR-Net outperforms all the competing methods on the DRIVE dataset, except for the \textit{specificity}, which is 0.5\% lower than that of the best performing IterNet \cite{li2020iternet}. It should be noted that EAR-Net outperforms the second best-performed method on DRIVE dataset in \textit{sensitivity}(SE) by a large margin (+2.3\%). Similarly, Highest \textit{sensitivity} is achieved by the proposed ERA-Net on the STARE dataset. The consistent superior performances regarding \textit{sensitivity} on both datasets indicate our method alleviates the problem of being dominated by the background pixels. For further clarification, we show segmentation results in Figures 2 and 6. Note that continuity losing problem may exist (see the red rectangulars in the Figures 2 and 6) in the ground truth from the STARE dataset due to the significant brightness. Nonetheless, even with the wrong supervision, ERA-Net still can learn the true blood vessel distribution. This can be observed by one randomly selected segmentation result shown in Figure 2, where these areas without continuity are connected fully or partly in the STARE dataset by EAR-Net. More segmentation results and examples are visualized in Figure 6.

\begin{figure*}[ht]
\begin{center}
\includegraphics[width=0.9\textwidth]{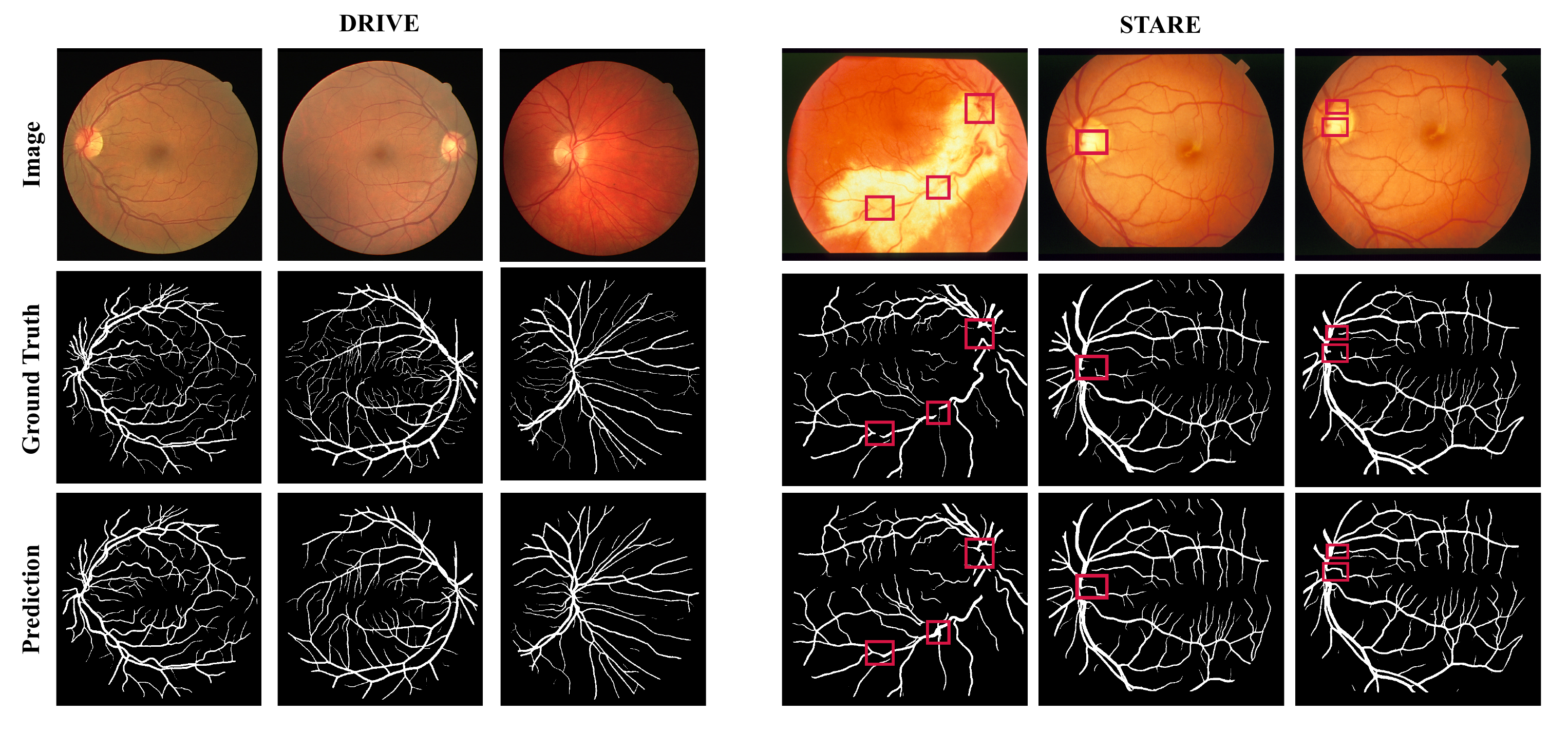}
\caption{Six randomly selected retinal blood segmentation results from the test set of DRIVE and STARE benchmarks by our proposed method.} \label{fig_more}
\end{center}
\end{figure*}

\begin{table}[!h]
\centering
\caption{Performance of different retinal blood vessels segmentation methods on the test set of DRIVE and all folds of the STARE.}
\begin{tabular}{c|ccc|ccc}
\toprule  
{\bfseries Methods} & \multicolumn{3}{c|}{ \bfseries DRIVE} &  \multicolumn{3}{c}{ \bfseries STARE} \\
\hline
 & ACC& SP& SE \ & \ ACC& SP& SE  \\
\hline
MBT \cite{sazak2019multiscale}&0.959&0.981&0.718&0.962&0.979&0.730 \\
HMM \cite{fan2018hierarchical}& 0.960&0.981&0.736&0.957&0.970&0.791 \\
LAD-OS \cite{zhang2016robust}& 0.947& 0.972&0.774&0.955&0.975&0.779 \\
IPACM \cite{zhao2015automated}& 0.954& 0.982&0.742&0.956&0.978&0.780\\
DeepVes \cite{fu2016deepvessel}&0.953&0.978&0.760&0.961&0.970&0.741 \\
UNet \cite{ronneberger2015u}&0.953&0.964&0.754&0.941&0.963&0.768 \\
WSF \cite{zhao2017automatic}&0.958&0.979&0.774&0.957&0.976&0.788 \\
R2-UNet \cite{ronneberger2015u}&0.956&0.981&0.779&0.971&0.986&0.830 \\
DUNet \cite{jin2019dunet,li2020iternet}&0.956&0.981&0.786&0.974&\bfseries{0.993}&0.681 \\
IterNet \cite{li2020iternet}&0.957&{\bfseries 0.983}&0.779&{\bfseries0.978}&0.992&0.772 \\
\bottomrule 
Ours&  {\bfseries 0.963}&0.978&{\bfseries 0.809}&0.969&0.980&{\bfseries0.840}\\
\bottomrule 
\end{tabular}
\end{table}

\begin{table*}[hb]
\caption{Ablation study of our EAR-Net on the test set of DRIVE and all folds of the STARE.}
\centering
\begin{tabular}{c|cccc|cccc}
\toprule  
{\bfseries Methods} & \multicolumn{4}{c|}{ \bfseries DRIVE} & \multicolumn{4}{c}{ \bfseries STARE} \\
\hline
 & ACC& SP& SE& mIoU \ & \ ACC& SP& SE& mIoU \\
\hline
Baseline& 0.9622&0.9780&0.7982&80.44&0.9684&0.9788&0.8410&81.75 \\
EAR-Net& 0.9630&0.9778&0.8088&80.86&0.9690&0.9795&0.8397&81.95 \\
\bottomrule 
\end{tabular}
\end{table*}
\subsection{Ablation Study}
We conduct an ablation study to verify the effectiveness of the proposed EAR-Net. An extra metric mean Intersection over Union m(IoU) is added to comprehensively evaluate the performance of the models. In can be observed in Table 2 that the EAR-Net is superior to the baseline HMSANet in most of evaluation metrics. We also note that there is a slight reduction on the \textit{specificity} on DRIVE (-0.02\%) and \textit{sensitivity} on STARE (-0.13\%). This is possibly due to that EAR-Net predicts the true distribution in the continuity losing areas, which may also lead to sensitivity reductions on STARE as these pixels are considered as the false positives with inaccurate ground truth. Noticeable improvement can be observed on the \textit{sensitivity} on the DRIVE dataset from 0.7982 to 0.8088 (+1.06\%), verifying the effectiveness of the proposed EAR-Net.

\section{Conclusion}
This paper has introduced a novel error attention refining network (ERA-Net) for effective retinal blood vessels segmentation. The error attention module has the function of predicting the possible errors during the refinement phase and driving the refinement focusing on these false predictions. This is achieved by regarding the differences between initial segmentation results and the ground truth as the ground truth to supervise the learning of the error attention maps. Through this way, the error attention module is less likely to trapped into the local optimum since it runs in a supervised manner. Experimental results on two common retinal datasets prove the superiority of our proposed method.

\bibliographystyle{IEEEtran}
\bibliography{ref.bib}

\end{document}